\documentclass{aa}
\usepackage{graphicx}
\newcommand{\harps}{{\textsc{harps}}}
\newcommand{\corot}{{\textsc{corot}}}
\newcommand{\most}{{\textsc{most}}}
\newcommand{\ind}[1]{_{\rm #1}}

\begin{document}
\title{Seismology and activity of the F type star
  HD\,49933~\thanks{Based on observations 
obtained with the \harps\ \'echelle spectrometer mounted on the 3.6-m telescope at ESO-La Silla Observatory (Chile) 
}}
\titlerunning{Oscillations and rotation modulation in HD 49933}
\author{
B. Mosser \inst{1}\and
F. Bouchy \inst{2}\and
C. Catala \inst{1}\and
E. Michel \inst{1}\and
R. Samadi \inst{1}\and
F. Th\'evenin \inst{3}\and
P. Eggenberger \inst{4}\and
D. Sosnowska\inst{4}\and
C. Moutou \inst{2}\and
A. Baglin \inst{1}}
\offprints{B. Mosser}

\institute{LESIA, CNRS UMR 8109,  Observatoire de Paris, 92195 Meudon cedex, France\\
\email{benoit.mosser@obspm.fr}
\and
LAM, CNRS UMR 6110, Traverse du Siphon BP8-13376 Marseille Cedex 12, France 
\and
Laboratoire Cassiop\'ee, CNRS UMR 6202, Observatoire de la C{\^o}te d'Azur,  BP 4229, 06304 Nice Cedex 4, France \and
Observatoire de Gen\`eve, 51 ch. des Maillettes, 1290 Sauverny, Switzerland
}
\date{Submitted October 18, 2004}

\abstract{
A 10-night asteroseismic observation programme has been conducted in January 2004 with the spectrometer \harps\ at the ES0 3.6-m telescope. 
The selected target, the 6th magnitude F5V star HD 49933, was chosen among the
prime candidates of \corot, 
the European space mission dedicated to characterize stellar oscillations mode with high precision photometry measurements. 
This star shows important line profiles variations, indicating a
  surprisingly high activity with respect to its low rotation rate. 
However, with the help of tools developed for disentangling the signatures of
activity and oscillations, 
we are able to observe its oscillation spectrum in the frequency range [1.2,
2.2 mHz]. 
We measure the large separation (88.7$\pm$0.4 $\mu$Hz) and the maximum
amplitude (around 0.4 $\pm$ 0.1 m\,s$^{-1}$ rms), respectively in agreement and 
marginal agreement with the  predicted values. 

}
\maketitle
\section{Introduction}

 F type stars lack seismic observations owing to the difficulty, 
due to the poor number of their spectral lines, to
get a high accuracy seismic signal with ground-based spectrometric measurements.
In this study, 
we present an asteroseismic observational programme conducted with the new generation
spectrometer \harps, 
installed at the ESO 3.6-m telescope in La Silla (Pepe et al. \cite{pepe02})
on a F type star which is a candidate target (HD 49933) of the space \corot\ mission
(Baglin et al.  \cite{baglin02}). The unique capabilities of the spectrograph \harps\ made possible to
measure, to identify and to characterize the
solar-like oscillations of such a 6th magnitude F star.
 
In this letter, we present first the observations (Sect.~2), then describe the
long term radial 
velocity pattern we observed (Sect.~3). The method we used in
order to correctly 
extract the radial velocities is exposed in Sect.~4. The seismic signature is confirmed and discussed in Sect.~5. Section~6 is devoted to conclusions. 


\section{Observations}

HD 49933 (HR 2530, ADS 5505A) was selected among the \corot\ prime targets.
This  F5\,V star lies in a region of the HR diagram not yet explored by
previous seismic observations. 
HD 49933 has an effective temperature of 6500 $\pm$ 100 K with an estimated
absolute magnitude $\rm M_V=3.39 \pm 0.10$. 
This star, with an iron abundance of $-0.37$ dex (Solano et al. \cite{solano04}), is sligthly metal poor compared to the Sun and to Procyon.
From observations of solar-like oscillations in 
Doppler velocity, Samadi et al. (\cite{samadi04}) find that the maximum in the 
oscillation velocity, $V\ind{max}$, scales approximatively as $(L/M)^{0.65}$ 
where $L$ and $M$ are the luminosity and the mass of the star respectively.
Such a scaling law then predicts $V\ind{max} \simeq 60 \pm 0.5$~cm\thinspace s$^{-1}$ rms for HD 49933.  
Spectrometric seismic observations being limited by the stellar line width, 
we had to select a slow rotator: HD 49933 had an estimated $v\sin i$ about 10.9 km\,s$^{-1}$   (Solano et al. \cite{solano04}),  
changed to 9.5$\pm$0.3 km\,s$^{-1}$ with our more precise measurements. 

\begin{figure}
\centering
\includegraphics[width=8.5cm]{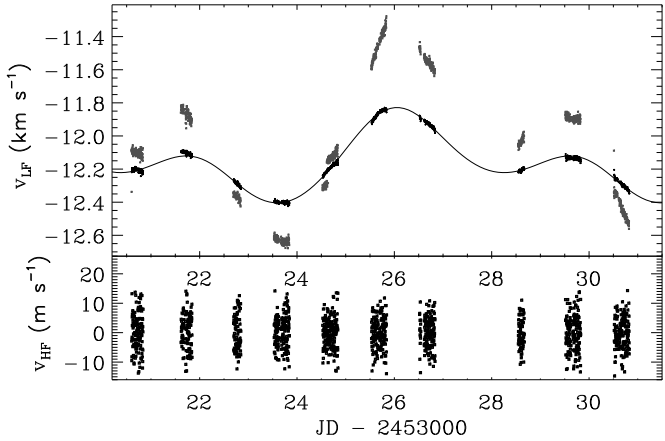}
\caption{Time series of the radial velocity of HD 49933 measured with the
  pipeline reduction of \harps. The average line bisector amplitudes, multiplied by the sign of the mean slope of the bisectors, are 
superimposed  (in grey) and translated for display purposes.  
The full signal (upper window) shows large variations, with a period about 7.9 days. 
The high frequency component (lower window) was corrected from the low frequency fit (thin 
solid line in the upper window) and also from the residual offsets and slopes.}
\label{timeseries}
\end{figure}

About 57 hours observations were recorded between January 15 and 25,
representing 1351 measurements. 
The exposure time was  120 s, giving one measurement each 153 s when including
overhead, and a Nyquist
frequency about 3.3 mHz above 
the predicted cutoff frequency at 2.2 mHz. 
The mean SNR in the spectrum (at 550 nm and with a spectral bin of 0.85 km\,s$^{-1}$) 
typically varies from 60 at high airmass to about 120. 
This SNR corresponds exactly to the contribution of the photon noise, which varies rapidly with the airmass, as the input flux in the entrance fiber. 
The mean observation duration did not exceed 7.5 hours per night at this
summer period.


\section{Long term radial velocity pattern, line asymmetries and activity}

The radial velocity curve of HD 49933, determined with the online pipeline reduction of
 \harps, shows large
 amplitude variations with a characteristic time scale of a few days
 (Fig.~\ref{timeseries}). 
Further measurements with the \textsc{coralie}  spectrometer on the Swiss 1.2-m telescope (Queloz et al. \cite{queloz00}) 
revealed that this long term radial velocity pattern is not stable over
 periods longer than 10 days,
 and cannot be attributed to one or several orbiting giant planets.

Suspecting that long term variations may be the signature of stellar activity,
 we examined the spectral line bisectors in our spectra. We first constructed average photospheric profiles, combining the profiles of the many
lines in the \harps\ spectra using the Least-Square Deconvolution (LSD) method described in Donati et al. (\cite{donati97}), the line pattern being constructed using a Kurucz model with $T\ind{eff}=6500$~K, $\log g$=4.5, and  [Fe/H]=$-0.3$. These parameters represent the closest choice to those
of HD 49933 in the model grids we have at our disposal. We have checked
that using adjacent models in the grid does not modify the results. Typical signal-to-noise ratios of 600 to 700 are obtained per velocity bin of 0.2 km\,s$^{-1}$ in the LSD average profiles.

\begin{figure}
\centering
\includegraphics[width=8.5cm]{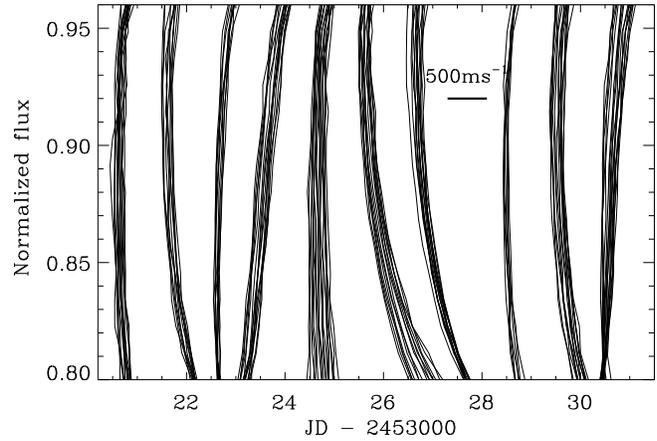}
\caption{Average line bisector amplitude of HD 49933.
The large variations identified in Fig.~\ref{timeseries} are related to the stellar activity  modulated by the rotation of the star. }
\label{bissec}
\end{figure}
The bisectors of the resulting average line profiles were then constructed as  the midpoints  of horizontal segments bounded by the sides of the profiles.
We find that the line bisectors of HD 49333 most often exhibit the classical
C-shape of cool stars, 
although with a larger distortion in the lower part than in the upper part of
the line, 
in contrast to most other observed cool star bisectors (Gray 1992). 
They have amplitudes up to about 1 km\,s$^{-1}$, measured as the velocity span
covered by the bisector, 
and they are highly variable.  The bisector 
variation is well visible on Fig.~\ref{bissec}, with the 8-day periodicity
suggested
 by the radial velocity curve in Fig.~\ref{timeseries}.
We conclude that this long-term radial velocity variation can be entirely 
attributed to line profile distortions, which are most likely related to
stellar surface  activity. 
The 8-day periodicity may be identified as rotational modulation of the
surface pattern giving rise to the variations. 
One probable interpretation is that the line distortions are due to the presence of one or several cool spots at the stellar surface, whose signature crosses the line profiles from blue to red during the star rotation, thus affecting the line bisectors. This interpretation is fully compatible with the apparent double-periodicity of the signal. The bumps produced by stellar spots are stronger near the bottom of the lines than in the line wings, which may explain the stronger 
variations of the line bisectors near their lower end.

Line bisector variations have already been reported in several cases, mostly for G and K type stars 
(L\'opez-Santiago et al. \cite{lopez03} and references therein), 
the bisector modulation being interpreted as the signature of one or several
cool starspots at the stellar surface. 
The observed bisector modulation amplitudes are usually small. In comparison
to these results, 
the larger amplitude (1 km\,s$^{-1}$) for the bisector modulation of HD 49933 is somewhat surprising for such a slow rotator 
($v\sin i=9.5$ km\,s$^{-1}$, $P\ind{rot} \simeq  8$ days). 
The detailed interpretation of these results and the modelling of the stellar surface structures will be proposed in a further paper.
 
\section{Disentangling Doppler shifts and line profile variations} 

\begin{figure}
\centering
\includegraphics[width=8.5cm]{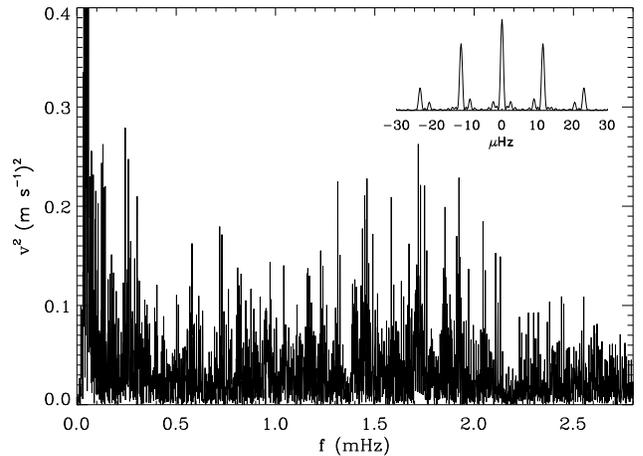}
\caption{Fourier spectrum of the velocity time series calculated with the  optimum method. 
The window function is shown in the inset. No high pass filtering was applied, except for the removal of the mean velocity slope of each night.}
\label{spectrum}
\end{figure}

The signature of activity occuring at low frequency should not affect the seismic signal at higher frequency. However, 
the large distortions of the lines related to activity may mimic
rapidly varying tiny Doppler shifts. 
Therefore, any method for extracting the radial velocity must be able to distinguish between a line distortion and a pure line shift.  
The numerical cross-correlation technique (Baranne et al.  \cite{baranne96}) 
of the reduction pipeline of \harps\  
uses as an input the theoretical position of the stellar lines.
The mean line profile is then reconstructed by interpolation of the observed lines corresponding to the theoretical positions. A Gaussian fit gives the Doppler shift, which is affected by  the line profile. Therefore, deformations of the line profile may give a spurious Doppler-like signal. 
The optimum method (Connes \cite{connes85}, Bouchy et al.  \cite{bouchy01}) is
in principle independent of the line profile.
Any spectral line is compared to itself at different times of
  observations and the wavelength shift is translated into a Doppler
 shift according to a first order expansion. 
The efficiency of the method requires a reference spectrum with a high
SNR. With the present set of observations, the method proves to be efficient 
in practive with a reference spectrum having a SNR greater than 500, resulting of the addition of more than 30 spectra. Therefore, the line profiles of the reference and of the spectra do not coincide, so that Doppler artefacts may be induced by lines distortions. 
Finally, both reduction methods give time series with similar
velocity noise of about 3.5 m\,s$^{-1}$ per 
integration time (Fig.~\ref{timeseries}). However, we cannot be sure that the signal in the Fourier spectrum plotted in Fig.~\ref{spectrum} is not appreciably contaminated by the stellar activity.

\begin{figure}
\centering
\includegraphics[width=8.5cm]{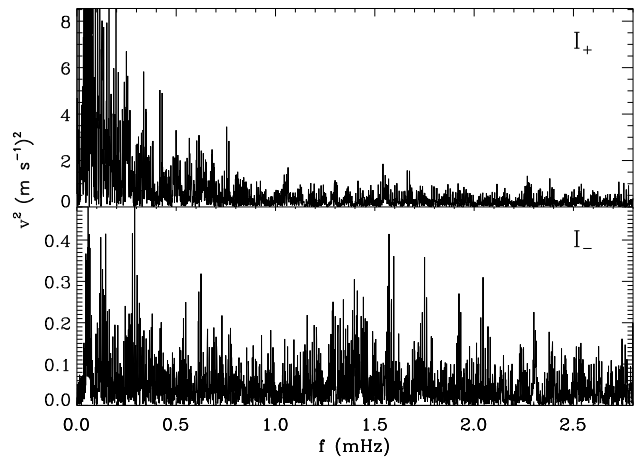}
\caption{Fourier spectrum of the time series $I_{+}$ (upper curve) and $I_{-}$
(lower curve), with a signal translated in velocity.  
The spectrum of $I_{+}$, representative of the deformation of the mean line,
does not show any excess power in the frequency range [1, 2.2 mHz], contrary
to the one of $I_{-}$, 
very similar to  Fig.~\ref{spectrum}.}
\label{figuresym}
\end{figure}

In order to disentangle the distortion and Doppler shift as precisely as
possible, we have analysed the LSD profile  by decomposing it in two terms. The mean line profile $I(t)$ being considered as a function of the velocity $v$, the intensities $I(t, v)$ and
$I(t, -v)$ explore it symmetrically, and give the integrated quantities:
$$
I_{+, -} (t) \ =\ \sum_0^V w(v) \ \bigl[I(t, v) \pm I(t, -v)\bigr]\\
$$
The normalized weighting function $w(v)$ is calculated from a Gaussian fit of
$I$, with $w$ proportional to the derivative of this fit, this enhancing the role of
the line wings comparatively to the continuum and the line center. The velocity $V$ is
 high enough for covering the whole profile. 
The spectrum of the symmetric component $I_{+} (t)$ shows no excess power in
the range [1.2, 2.2 mHz] as seen on Fig.~\ref{figuresym}, contrary to the antisymmetric signal
 $I_{-} (t)$, which is very similar to the spectrum of Fig.~\ref{spectrum}:
$I_{+}$ is sensitive to the line broadening and distortion, while $I_{-}$ is not, but sensitive to a line shift. This treatment is then able to extract the 
information of a pure 
Doppler signal, and confirms that the observed Fourier spectrum  is not affected by distortions.

\section{The seismic signature}

\begin{figure}
\centering
\includegraphics[width=8.cm]{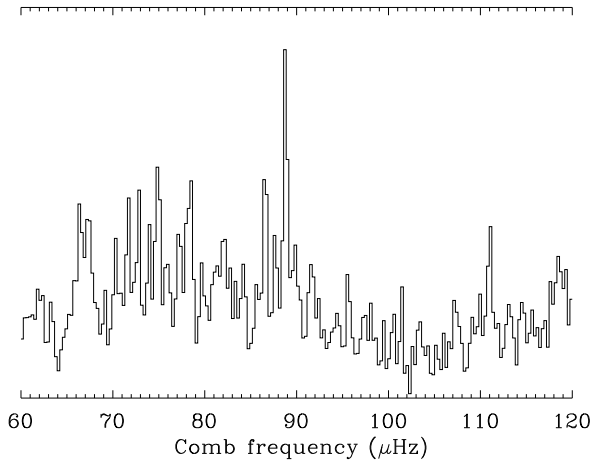}
\caption{Comb response of the spectrum of $I_{-}$
between 1.2 and 2.2 mHz. The signal around 88.7 $\mu$Hz is the signature of low degree solar-like pressure modes.} 
\label{comb}
\end{figure}

The Fourier spectrum shows an excess power in the frequency range [1.2,
  2.2 mHz],
peaking at 1.7 mHz near the predicted value of 1.3 mHz derived from the scaling law Bedding \& Kjeldsen (\cite{bedding03}). 
The comb analysis (Mosser et al. \cite{mosser98}) exhibits a clear signature 
around $\Delta\nu = 88.7\pm 0.4~\mu$Hz (Fig.~\ref{comb}), representative of the seismic large separation.
We computed evolutionary models for HD 49933, using the CESAM code
 (Morel \cite{morel97}).
We considered the global parameters of HD 49933 (6500 K, [Fe/H] = $-0.37$,  $M\ind{V} = 3.39$)
and classical physical options for this kind of objects. 
Heavy element abundance Z$= 7\ 10^{-3}$ is deduced from [Fe/H] and we adopted for helium Y=0.255 with 
Y$\ind{primordial}$=0.240 at Z$\ind{primordial}$=0,with Z/X$_\odot$ = 0.0245 and an enrichment galactic law dY/dZ=2. The low metallicity induces a  small mass for this range of temperature,  so that the preferred mass is around 1.15 $M_\odot$, with an age between 3.3 and 4.0 Gyr. For all computed models,
the associated seismic large separation is found to vary between 75 and 88 $\mu$Hz. The 88.7 $\mu$Hz spacing found in the data is thus consistent with this range of expected values, and is in favor of a young age.  

In order to estimate the maximum oscillation amplitude, we have constructed  synthetic time series, based on a theoretical low degree p-modes eigenfrequency pattern. The modes lifetimes are estimated from the eigenfrequency widths, between 1 and 4 $\mu$Hz FWHM (Houdek et  al. \cite{houdek99}). The maximum amplitudes are assumed to follow a Gaussian distribution in frequency. The synthetic power spectra are then calculated 
using the model of a stochastically excited, damped harmonic oscillator (Anderson et al. \cite{anderson90}).
We assume that the spectrum is composed of this synthetic signal and a white noise. The maximum amplitude was then determined in order to obtain comparable energy per frequency bin in the synthetic and observed spectra. A Monte-Carlo approach finally shows that the best agreement is for a signal maximum amplitude about 0.4$\pm$0.1 m\,s$^{-1}$ rms, with a Gaussian envelope centered at 1.7 mHz and with a 1.0 mHz FWHM. This estimate is not much sensitive to the assumed eigenfrequency widths: an uncertainty of $\pm$50\% of their values adds an uncertainty smaller than 15\%. 
The agreement with the predicted scaling (60$\pm$5 cm\,s$^{-1}$ rms) appears to be marginal.

\section{Conclusion}

We have been able to measure with \harps\ the p-modes oscillation amplitude  and the large separation of HD 49933. 
This confirms that ground-based asteroseismologic observations with
\harps\ provide excellent solar-like oscillations studies, even for a faint ($m\ind{v}\simeq 5.7$) F type star with a $v\sin i $ near 10 km\,s$^{-1}$ and exhibiting activity. 
For a star such as HD 49933, ground-based spectrometric measurements are less efficient  than for cooler types stars, with more and narrower lines. 
However, we have demonstrated that spectrometric measurements are the
  appropriate tools for disentangling  the signatures of oscillations and activity.
This problem is particularly important for intermediate mass stars, as most of
the \corot\ targets, that contain information on the extension of the core
 and on the angular momentum transfer in the early phases of their
 lives.
The question of contamination of the seismic signal by activity in the p-mode
frequency range has been raised by Matthews (\cite{matthews04}) as a possible 
cause of the non detection of oscillations on Procyon with \most. Our
observations allow an 
unambiguous distinction between the stellar activity and the seismic signal.

\begin{acknowledgements}

We thank the Gen\`eve group of \harps\ for his precious help before and during
the run. 
We thank Christophe Lovis for the follow up measurements conducted with
\textsc{coralie} 
at the swiss 1.2-m telescope, Pierre Magain and Michael Gillon for modelling the stellar spectrum. We also thank the anonymous referee for his constructive comments.

\end{acknowledgements}

\end{document}